\documentclass[aps,prl,reprint,superscriptaddress]{revtex4-1}
\usepackage{amsfonts}
\usepackage{amssymb}
\usepackage{amsmath}
\usepackage{graphicx}
\usepackage{epsfig}
\usepackage{array}
\usepackage{braket}
\usepackage{multirow}
\usepackage[table,xcdraw]{xcolor}

\begin{document}

\title{Anomalous Bloch oscillation and electrical switching of edge
magnetization in bilayer graphene nanoribbon}
\author{Tixuan Tan}
\author{Fengren Fan}
\author{Ci Li}
\email{oldsmith@hku.hk}
\author{Wang Yao}
\email{wangyao@hku.hk}
\affiliation{Department of Physics, The University of Hong Kong, Hong Kong,
China}
\affiliation{HKU-UCAS Joint Institute of Theoretical and
Computational Physics at Hong Kong, China}
\begin{abstract}
Graphene features topological edge bands that connect the pair of Dirac points through either sectors of the 1D Brillouin zone depending on edge configurations (zigzag or bearded). Because of their flat dispersion, spontaneous edge magnetisation can arise from Coulomb interaction in graphene nanoribbons, which has caught remarkable interest. We find an anomalous Bloch oscillation in such edge bands, in which the flat dispersion freezes electron motion along the field direction, while the topological connection of the bands through the bulk leads to electron oscillation in the transverse direction between edges of different configurations on opposite sides/layers of a bilayer ribbon. Our Hubbard-model mean-field calculation shows that this phenomenon can be exploited for electrical switching of edge magnetisation configurations.
\end{abstract}

\maketitle


\textit{Introduction.} The existence and behavior of edge states are always attractive in the study of solid physics due to their distinct properties in contrast to bulk states. For the monolayer graphene (MLG) that has
zero-gap band structure where the conduction and valance bands touch at the Dirac points \cite{Neto,Dres,Wal,Gus,Novo2,Zhang}, edge states in MLG ribbon appear
as flat bands at the Fermi level, connecting the bulk Dirac points through either sectors of the 1D Brillouin zone depending on edge configurations (zigzag or bearded) \cite{Neto,Dres,Nak,Ryu,Bre,Yao1,Del}.
When the bulk gap is opened,
these flat-band edge states can be continuously tuned into gapless chiral edge modes through bias control on the edge \cite{Yao1},
which have similar origin to the topological domain wall modes in bilayer graphene (BLG) \cite{Mar,Zar,Ju}.
With the non-trivial topological properties and relation to the bulk valley transport \cite{Yao1,Yao}, these chiral modes have also been explored in other context such as the laser-induced Floquet system\cite{Per} and gapped nanomechanical graphene \cite{Xi}.

\begin{figure}[tbp]
\begin{center}
\includegraphics[width=0.48\textwidth]{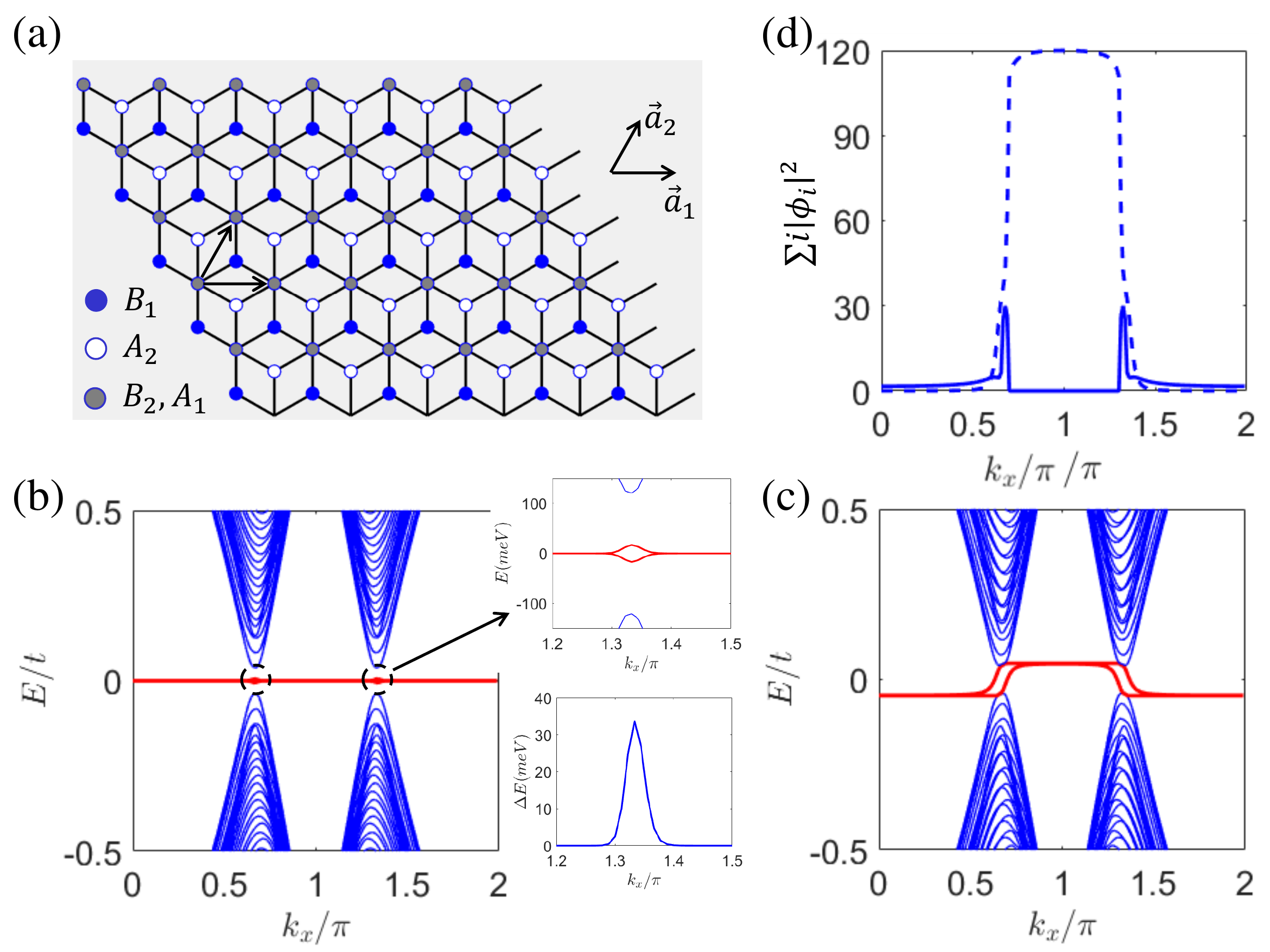}
\end{center}
\caption{(Color online) The schematic illustration of the atomic structure
of biased bea-zig BLG ribbons and related bulk chiral modes. The total
number of unit cell in finite direction of the ribbon is chosen as $N=60$,
i.e., $120$ atoms for each layer for all plots. (a) The atomic structure of
biased bea-zig BLG ribbons. The unit vectors are denoted by $\vec{a}_{1}$
and $\vec{a}_{2}$. Ribbons are assumed to be infinite along $x$ ,viz. $\vec{a}_{1}$, direction.
(b) The band structure for the bea-zig BLG ribbon with no bias. Two bands of
edge states are highlighted by the solid red lines. The zoomed-in band
structure in the black dashed circle is shown next to (b) and the energy difference between two edge bands (red solid lines) is presented below. (c) The band structure for the biased
bea-zig BLG ribbon. Two bands of edge states are highlighted by the solid
red lines. (d) The wave function distributions of bulk chiral modes, where
the wave function is in the form $|\protect\varphi \rangle =\sum \protect%
\phi _{i}|i\rangle $ with $i$ labelling sites along $\vec{a}_{2}$ direction.
This wave function depicts the higher of the two red bands, with
dashed(solid) line referring to the distribution on top(bottom) layer. The
parameters are chosen as the interlayer bias $U_{1}=0.1t$ for (b) and $%
U_{1}=0$ for (c).  $t=3.16\mathrm{eV}$ for all
plots.}
\label{fig1}
\end{figure}
\vspace{-0.06cm}

The flat edge band, on the other hand, promises the emergence of magnetism when electron interaction is taken into account~\cite{Son,Roj,Slo,Yaz}.
In zigzag MLG nano-ribbons, the repulsive on-site Coulomb interaction is shown to introduce ground state spin polarization (SP)
on the edges, which can be either antiferromagnetic (AFM) or ferromagnetic (FM), i.e., the localized magnetic
moments at the opposite edges of the ribbon is antiparallel or parallel \cite%
{Son,Roj,Slo}, turning the system into a semiconductor or a
conductor (metal), respectively \cite{Yaz,Son}. Based on this phenomenon,
some interesting applications in spintronics have been proposed such as
half-metallicity induced by in-plane electric field \cite{Son,Dut} and
control of the spin transport by introducing defects \cite{Wim}. Similar
magnetic effects have also been found in the non-standard-shaped MLG ribbons
\cite{Wan,Li}.
Compared with MLG ribbon, magnetism in BLG materials and nanostructures is
less studied. Most of earlier works focused on the half-metallicity and related magnetic
effects in the zigzag BLG nanoribbons \cite%
{Sah,Kot,Lee,Sza} or bulk BLG system \cite{Yuan}.

In this letter, we focus on the motions of electrons correlated with the edge states in
bea-zig BLG (top and bottom layer of a BLG ribbon have zigzag and bearded edges
respectively, c.f. Fig. \ref{fig1}(c)), which host flat-dispersion edge bands in the
entire 1D Brillouin zone \cite{Ti}. Under interlayer bias, we find that gapless chiral modes appear in the ribbon bulk near the Dirac
point, connecting states localised on opposite edges and layers. Bloch oscillation in the edge bands
driven by electric field along the ribbon has an unusual form in the real space, where the electrons predominantly oscillate in the transverse direction between opposite edges and layers.
Such phenomenon represents an interesting aspect of Bloch oscillation in momentum space, reflected in real space due to the spatial character of topological edge bands.
With Hubbard interaction included through a mean-field approach, we show this anomalous Bloch oscillation can be exploited for electric field control of transition between different edge magnetic states of the BLG nano-ribbon, which points to a new possibility of spintronic control.

\begin{figure*}[tbp]
\begin{center}
\includegraphics[width=0.8\textwidth]{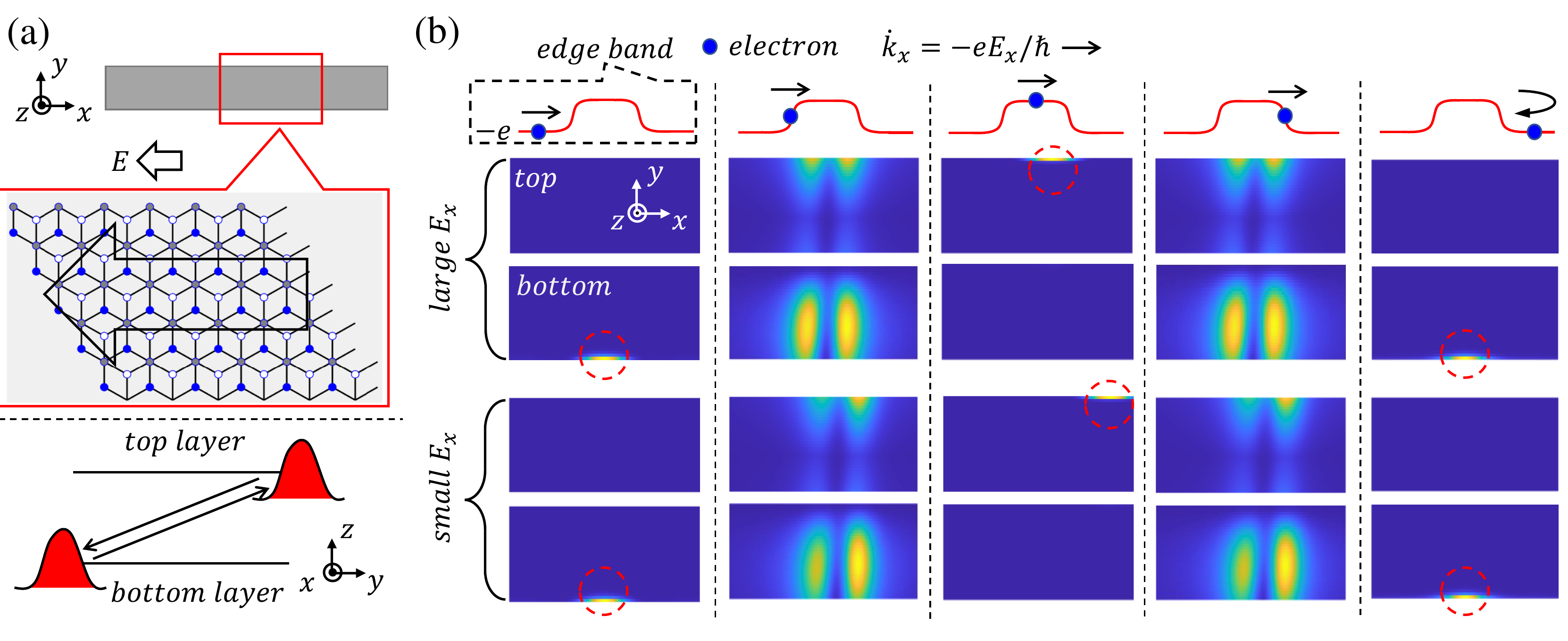}
\end{center}
\caption{(Color online) The numerical simulation of the motion of electrons
in real space based on the edge states of the biased bea-zig BLG. The
dynamics of electrons is simulated by using a Gaussian wave packets in the
momentum space, following the procedure outlined in Ref. \protect\cite%
{Hartmann}. (a) Top panel: A schematic figure of the BLG ribbon on which the
simulation is conducted. $x$ direction is the infinite (periodic direction),
while $y$ is the finite direction. It is chosen that the length(width) of
the ribbon ($x(y)$ direction) is $4800a$ and $48a$, with $a$ being the
lattice constant of graphene. The red box circles the plot range of the
simulation, which covers a length of $800a$. Bottom panel: Schematic
illustration of the movement of electrons perpendicular to the applied
external electric field $E_x$ in the real space, which is the side view in
the $yz$ plane. (b) The result of the numerical simulation when the electron
on the edge band moves to different position in $k_x$ space, shown in the
top row. The color represents the magnitude of wave function in real space,
increasing from blue to yellow. The parameters used is $E_x=-1.7%
\times10^{-2}/-2.8\times10^{-3} \mathrm{V/nm}$ for the middle and bottom
row, respectively. The color scale is matched within each set of graphs
(two), but may vary from set to set.}
\label{fig2}
\end{figure*}
\vspace{-0.06cm}

\textit{Anomalous Bloch Oscillation}
The electronic properties of the
biased bea-zig BLG can be described by the tight-binding Hamiltonian \cite%
{Ti}%
\begin{equation}
H_{\mathrm{BLG}}=H_{\mathrm{bea}}^{1}+H_{\mathrm{zig}}^{2}+H_{\mathrm{int}%
}+H_{\mathrm{bias}},
\end{equation}%
where%
\begin{equation}
H_{\mathrm{bea/zig}}^{l}=-t\sum_{\left\langle i,j\right\rangle ,\sigma
}\left( c_{l,i,\sigma }^{\dagger }c_{l,j,\sigma }+\mathrm{H.c.}\right)
\end{equation}%
represents the tight-binding Hamiltonian of MLG with bearded/zigzag edges
and $l=1,2$ are labels of the bottom and top layers respectively. $t$
denotes the nearest-neighbor (NN) hopping in MLG with $c_{i,\sigma
}^{\dagger }(c_{i,\sigma })$ being the creation (annihilation) operator of $%
\sigma $-spin electron on site $i$ in the ribbon, and $\sum_{\left\langle
i,j\right\rangle ,\sigma }$ only sums over NN pairs. Since there is no
Hubbard interaction, i.e., no interaction between different spins, the index
$\sigma $ can be ignored.
\begin{equation}
H_{\mathrm{bias}}=\sum_{l,i,\sigma }\left( -1\right) ^{l}\frac{U_{1}}{2}%
\left( c_{l,i,\sigma }^{\dagger }c_{l,i,\sigma }+\mathrm{H.c.}\right)
\end{equation}%
refers to the interlayer bias $U_{1}$.  The van der Waals interaction between two layers \cite{Neto,Mc}
is described by $H_{\mathrm{int}}$. In this paper we only consider the NN
interlayer coupling $\gamma _{1}$ for simplicity. Here, we take $t=3.16%
\mathrm{eV}$, $\gamma _{1}=0.381\mathrm{eV}$ as typical experimental values
for AB-stacked BLGs \cite{Kuz}.

We take $x$ direction as the infinite direction of the BLG ribbon. For this
specific structure, it has been shown in our earlier work \cite{Ti} that
there are two degenerate non-dispersive edge bands \vspace{-0.06cm} in the
whole $k_{x}$ region when $U_{1}=0$. They are related with an
interlayer-coupling-protected topological phase transition between two
non-trivial topological phases characterized by winding number $W=-1$ and $%
W=1$ when crossing the Dirac point. When $U_{1}\neq 0$, topological edge
states still exist in the whole $k_{x}$ region with different energy when
crossing the Dirac point. They are connected by a pair of bulk chiral modes,
as shown in Fig. \ref{fig1}(c). The degeneracy between edges are lifted due
to the broken chiral symmetry, as discussed in \cite{Ti}. The bulk chiral
modes connect states residing in opposite edges of different layers. The
connection must be done through the bulk, hence the name. The main
difference between the bulk chiral mode we discuss here and those edge
chiral modes discussed in previous literatures \cite{Yao1,Mar,Per,Zar,Ju}
is obvious: bulk chiral modes are obtained by imposing a bias on the bulk,
and the wave function of the chiral modes are also distributed in the bulk.
while edge chiral modes are obtained via tuning edge on-site energy, whose
wave function is localized at the edge. The fact that these connecting
chiral modes are bulk states are by themselves interesting in terms of
transport. It is often the case the bulk part of the material is insulator
so that all electrical transport are dependent on topologically protected
edge states. However, in this system we discuss, transport is done by  two
special pairs of chiral bulk states, one at each valley. These special bulk
states at the Fermi level are present only because that there are edge
states populating different area of the momentum space, which can not be
found in the bulk spectrum of the BLG as shown by the blue bands in Fig. \ref%
{fig1}(c).

An interesting application of these bulk chiral modes is Bloch oscillation in topological insulators \cite{Hartmann,Viktor,Liu,Bloch,Chunyan,Lib}.
When applying an electric field $eE_{x}=\frac{\partial A(t)}{\partial t}$
along the infinite direction ($x$ direction in our assumption) of the
ribbon, the motion of electrons in the edge states can be approximately
described by semiclassical equations of a wave packet. It indicates that the
wave vector of the electron will evolve according to $\dot{k}_{x}=-
eE_{x}/\hbar $, as shown in Fig. \ref{fig2}(a). Because of the existence of the bulk chiral
modes connecting two opposite edges of two different layers, the transition
of electrons from left edge of one layer to the right edge of the other
layer is possible, giving rise to a Hall-effect-like behaviour of electrons,
as illustrated in Fig. \ref{fig2}(a).

The naive conjecture as above may be undermined by the smallness of the gap
opened by the bias. Numerically, a bias as large as $U_{1}=300\mathrm{meV}$
can open a gap of around $5\mathrm{meV}$ between two edge bands in the flat
part. Landau-Zener tunneling may cause the electrons to transit between
different edge bands, breaking down our conjecture based upon single band
picture. However, such a transition is suppressed by the fact that there is
no spatial overlap between the wavefunctions of two flat edge bands, i.e. $\braket{lower|H_{E_x}|upper}\approx 0$ \cite%
{Note4}. Here, we use a wave packet to simulate the motion of electrons in
real space. It shows the expected trajectory, which is shown in Fig. \ref%
{fig2}(b). Notice that in a full cycle, i.e., the momentum $k_{x}$ evolves
from $0$ to $2\pi $, a wave packet will move both in x direction and in y
direction. However, the motion in these two directions are different. Since
only the part of edge bands near the Dirac points (bulk chiral modes) has
non-vanishing group velocity and the time of wave packet staying in this
region is inversely proportional to the field strength, the range of motion
of the wave packet in the $x$ direction is inversely proportional to
electric field strength $E_{x}$, which can be observed in Fig. \ref{fig2}%
(b). Also, this range is proportional to the bias as it determines how dispersive the chiral modes are. However, the range of
motion of the anomalous oscillation in the transverse direction, i.e., $y$
direction, is independent of the field strength, which must be the whole
width of the ribbon. Here we have used $U_{1}=300\mathrm{meV}$ in numerical
simulation to make bulk chiral modes in Fig. \ref{fig1} more obvious. Similar
phenomena can still be observed for smaller $U_{1}$ , or even for $U_{1}\approx 0$, where the motion in the $x$ direction is frozen, leaving the
oscillation in the transverse direction unaffected \cite{Supp}. It should be
noted that this phenomena is unique to this bea-zig BLG ribbon, as there are
edge bands detached from the bulk bands in the whole Brillouin zone. It is
not possible to observe it in the BLG zigzag-zigzag ribbon, as all four of
its edge bands only exist in certain region in momentum space.

\begin{figure*}[tbp]
\begin{center}
\includegraphics[width=0.8\textwidth]{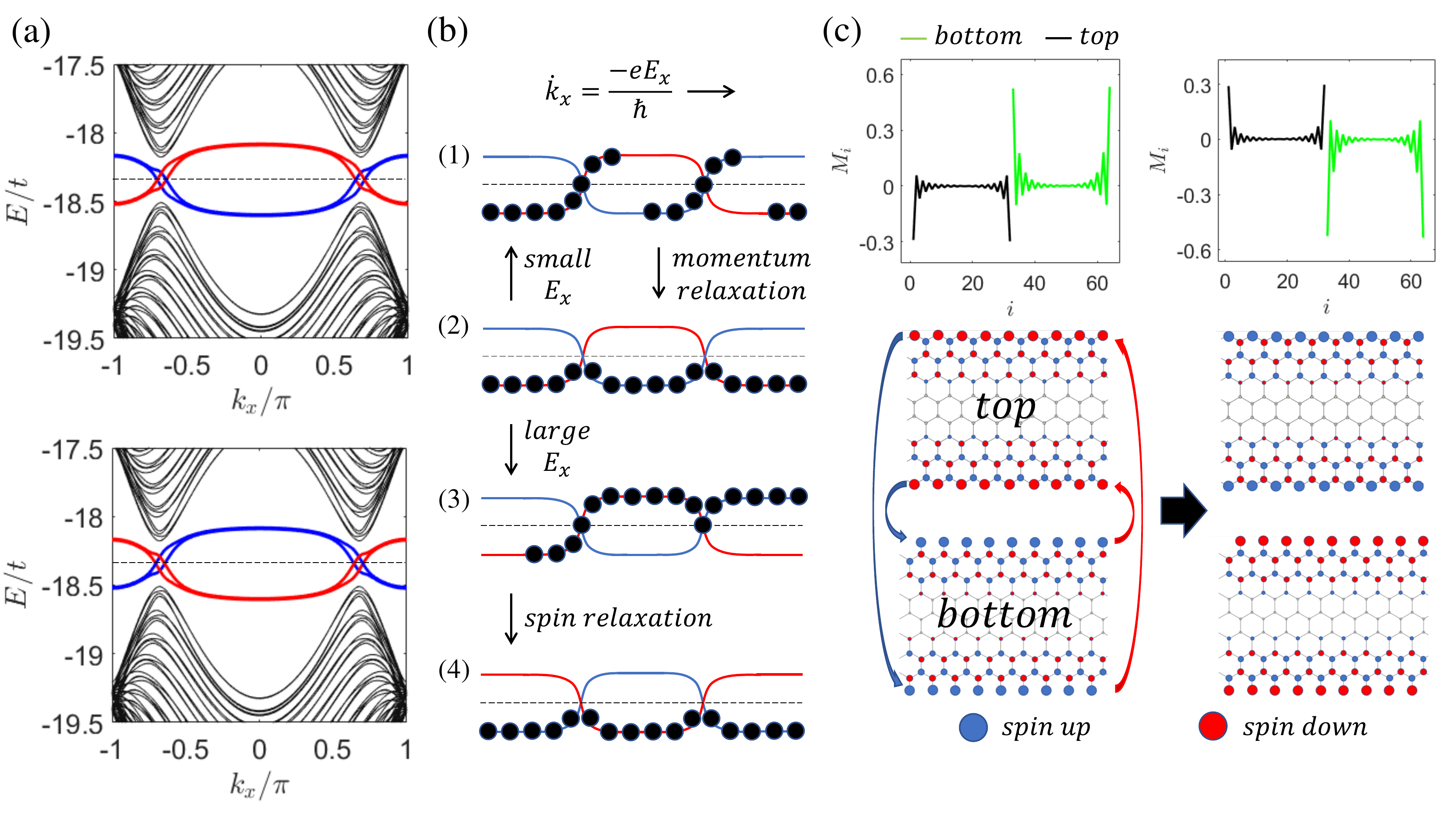}
\end{center}
\caption{(Color online)\textbf{(a)} Band structure of two self-consistent ground
states of bea-zig BLG ribbon in Hubbard model, corresponding to Conf. $4$ in
the Table. \protect\ref{TableI}. Black dashed line is the Fermi level. Black
solid lines are bulk bands . Blue(Red)
solid lines are edge bands of spin up(down) electrons. \textbf{(b)} Schematic
illustration of the Bloch oscillation in the same system. (2)((4))
corresponds to the top(bottom) panel of (a). (1)((3)) corresponds to the
situation where the deviation from the initial state (top panel of (a)) is
small(large). The black circle are electrons occupying bands. \textbf{(c)} Top
panel: SP configurations corresponding to (a) (top(bottom) panel of (a)
corresponds to left(right), respectively), which is defined by the magnetic
moment $M_i$ \protect\cite{Note2}. The blue(orange) solid line represents
situation on the top(bottom) layer. The horizontal axis is the site index,
increasing along the finite direction of the ribbon. The top layer atoms
constitute the first half while the bottom layer atoms constitute the second
half. Bottom panel: The schematic illustration of the SP distribution in
real space, corresponding to SP configurations shown in the top panel. A
transition between these two configurations can be induced by the process
illustrated in (b).}
\label{fig3}
\end{figure*}
\vspace{-0.06cm}

\textit{Electrical switching of edge magnetisation configurations}
Magnetic effects of the edge states in the BLG ribbon can be described by
self-consistent mean-field calculation when adding the Hubbard interaction $%
H_{U}$ to the usual tight-binding Hamiltonian $H_{\mathrm{BLG}}$ of the BLG
ribbon \cite{Yaz}
\begin{equation}
H=H_{\mathrm{BLG}}+H_{U},H_{U}=U\sum_{i}n_{i,\uparrow }n_{\imath ,\downarrow
}.
\end{equation}%
The Hubbard interaction $H_{U}$ represents the electron-electron interaction
in the form of the repulsive on-site Coulomb interaction, where $%
n_{i,\uparrow (\downarrow )}\equiv a_{i,\uparrow (\downarrow )}^{\dagger
}a_{i,\uparrow (\downarrow )}$ is the occupation number operator and $U>0$
describes its magnitude \cite{Yaz}. In this letter, we take $U=1.2t $ \cite%
{Yaz}.

\begin{table}[tbph]
\caption{Summary of self-consistent solutions of bea-zig BLG ribbons. $\pm $
means the sign of the magnetic moment at the corresponding edge. tu(td)
represents the up(down) edge of the top layer, while bu(bd) are for those of
the bottom layer. The up(down) is relative to the $y$ direction as shown in
Fig. \protect\ref{fig1}. $SC$ and $C$ denote semiconductor and metal
(conductor), respectively. There are altogether $16$ configurations, where
the other $8$ configurations are obtained from these by an exchange between
spin-up electrons and spin-down electrons.}
\label{TableI}
\begin{center}
\begin{tabular}{|c|c|c|c|c|c|}
\hline
& tu & td & bu & bd & SC/C \\ \hline
Conf. 1 & $+$ & $+$ & $+$ & $+$ & $C$ \\ \hline
Conf. 2 & $+$ & $+$ & $+$ & $-$ & $C$ \\ \hline
Conf. 3 & $+$ & $+$ & $-$ & $+$ & $C$ \\ \hline
Conf. 4 & $+$ & $+$ & $-$ & $-$ & $C$ \\ \hline
Conf. 5 & $+$ & $-$ & $+$ & $+$ & $C$ \\ \hline
Conf. 6 & $+$ & $-$ & $+$ & $-$ & $SC$ \\ \hline
Conf. 7 & $+$ & $-$ & $-$ & $+$ & $SC$ \\ \hline
Conf. 8 & $-$ & $+$ & $+$ & $+$ & $C$ \\ \hline
\end{tabular}%
\end{center}
\end{table}

Here, since the interaction between different spins is no longer zero, the
spin degrees of freedom $\sigma $ should be considered. We show the band
structure of $H$ for both spins in Fig. \ref{fig4}(a). The calculating
detail and a brief review of the model can be found in the Supplementary
\cite{Supp}. Similar bulk chiral modes connecting edge modes appear as those arising in the non-hubbard spinless bea-zig BLG ribbon, with opposite chirality for two spins, as shown in Fig. \ref{fig4}(a). The SP configurations are not limited to two simple types (AFM/FM)\vspace{-0.06cm} as found in the MLG ribbon in
Ref. \cite{Yaz}, there are $8$ inequivalent types of SP for BLG cases, as shown
in the Table. \ref{TableI}. Here we only demonstrate two types of band
structures corresponding to configuration 4 in Table. \ref{TableI} for
simplicity in Fig. \ref{fig4}(a).
\begin{figure}[tbp]
\begin{center}
\includegraphics[width=0.48\textwidth]{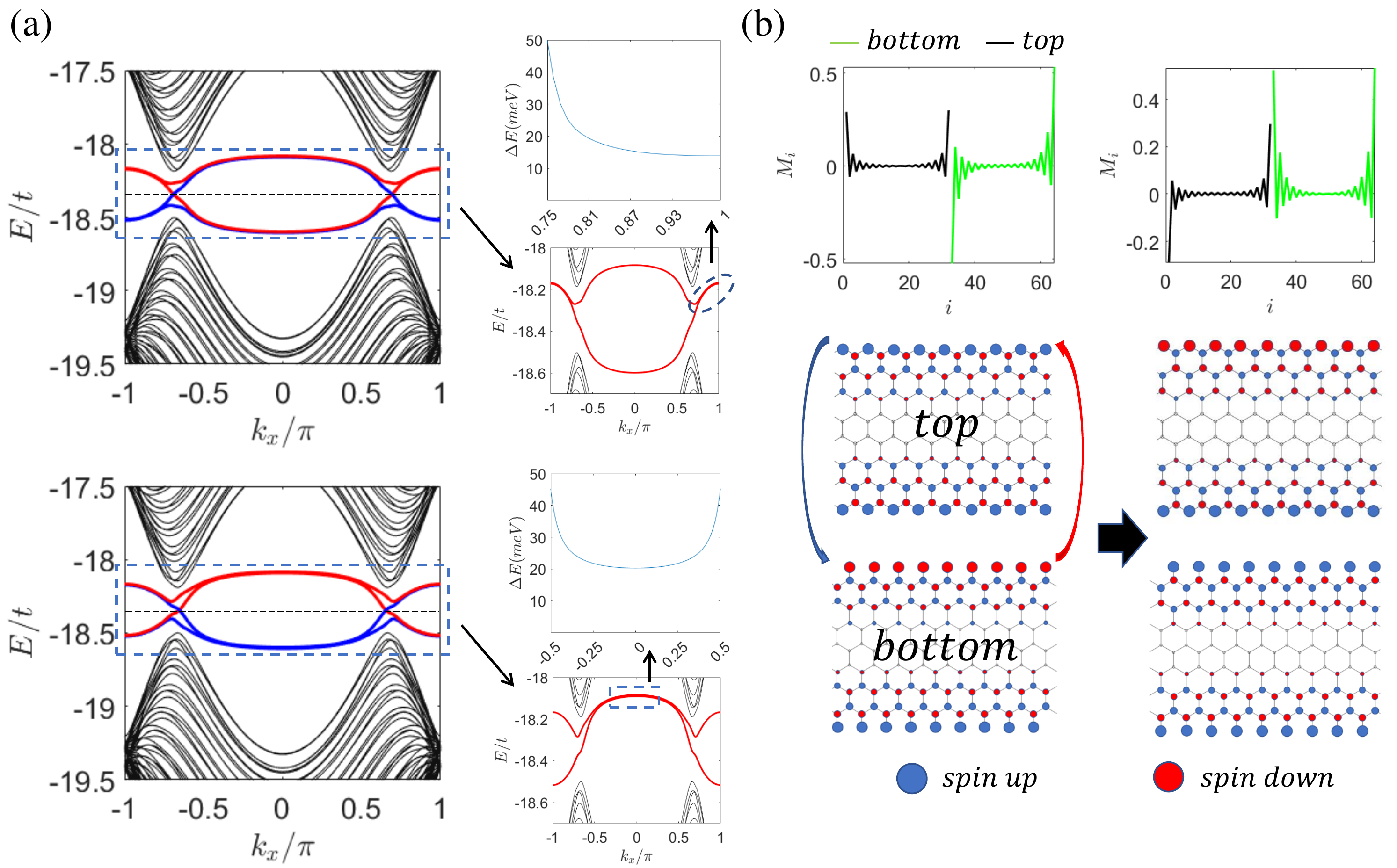}
\end{center}
\caption{(Color online) (a) Band structure of two self-consistent ground
states of bea-zig BLG ribbon in Hubbard model, corresponding to Conf. $3$
and $8$ in the Table. \protect\ref{TableI}, respectively. Black dashed line
is the Fermi level. Black solid lines are bulk bands. Blue(Red) solid lines are edge bands of spin up(down)
electrons. The zoomed-in band structures in the blue dashed box are shown next to it and related energy difference (around $20\mathrm{meV}$) between the nearly degenerate part is presented above each zoomed-in band structure. (b)
Top panel: SP configurations corresponding to (a) (top(bottom) panel of (a)
corresponds to left(right)), which is defined by the magnetic moment $M_{i}$
\protect\cite{Note2}. Bottom panel: The schematic illustration of the SP
distribution in real space, corresponding to two SP configurations shown in
the top panel. A transition between these two SP configurations can be
induced based on the process similar to that shown in Fig. \protect\ref{fig3}%
(b).}
\label{fig4}
\end{figure}
\vspace{-0.06cm}

 The essence of the result in the spinless model is that an external
field $E_x$ can push electrons through the momentum space, thus causing
corresponding motion in the real space. In a spinless model, the occupation
of the electrons in the momentum space will not affect the band structure,
while the band structure of the Hubbard model is dependent on the momentum
space distribution of electrons \cite{Yaz,Supp}. Thus, not all
configurations in the momentum space is a self-consistent ground state at
equilibrium. In fact, as we discussed in last section, there are only $%
16(8\times2)$ self-consistent solutions of the model. Any unstable
configurations should relax to one of them. An electric field $E_x$ will
push the system out of equilibrium by shifting the position of the electrons
in momentum space. If the shift is small, it is reasonable to expect that
the system should relax to the initial condition. However, if the shift is
large enough in momentum space \cite{Note3}, it is possible that the system
relax to another ground state when trying to reach equilibrium. The idea of
this procedure is schematically illustrated in Fig. \ref{fig3}(b).

Numerically, this is verified by giving different initial guess to the
electron's momentum space distribution, which would stably converge to
different ground state, as illustrated in Fig. \ref{fig3}(a) and (b).
Initially, spin up electrons occupy the region $k_x\in [-\frac{2\pi}{3},%
\frac{2\pi}{3}]$ while spin down electrons occupy the region $k_x\in[-\pi,-%
\frac{2\pi}{3}]\cup[\frac{2\pi}{3},\pi]$. If the occupation is pushed to the
positive $k_x$ direction by a small distance, it will stably converge to the
initial state, i.e. top panel of Fig. \ref{fig3}(a). If electrons are pushed
away more from their initial state, it would stably converge to an
equilibrium state that differs from its initial state by an exchange of
spin, i.e. bottom panel of Fig. \ref{fig3}(a). The corresponding SP
configuration is given in Fig. \ref{fig3}(c). This is an intra-configuration
transition between ground states, i.e., Conf. $4$ of Table. \ref{TableI}.

Moreover, we found that the inter-configuration transition between ground
states is possible by following the similar process illustrated in Fig. \ref%
{fig3}. This is shown in Fig. \ref{fig4}, where Conf. $3$ in Table. \ref%
{TableI} can be transformed into Conf. $8$ through the exchange of spins
between the same edge of different layers. The switch of edge-magnetization only
happens between two partially filled bands since there are always finite
gaps (around $20\mathrm{meV}$) between these two bands and the other two bands,
i.e., one empty edge band and one fully filled edge band, as shown in Fig. \ref%
{fig4}(a).

\textit{Conclusions and discussions.} In summary, we study the motions of
electrons related with the edge states in bea-zig BLG with and without
Hubbard interaction. The chiral modes present in the system without Hubbard
interaction is no longer edge chiral states appearing in the spinless MLG
ribbons but bulk states. These bulk chiral modes are unconventional, which
connect two non-trivial topological phases and can be modulated by a
mechanism adjusting property of the bulk instead of adjustments on the edge.
When applying an electric field along the infinite direction of the ribbon,
the motion of electrons in the edge states can be approximately described by
Bloch oscillation based on semiclassical equations of wavepackets.
This leads to an anomalous transverse oscillation since the bulk chiral
modes connect opposite edges of different layers. For the same system with
the Hubbard interaction, the Hall-effect-like behavior persists when the
electric field is applied, protected by the spatial character of the
topological edge bands. Besides, we also exploit the possibility of electrical switching of edge magnetisation in Hubbard model using this bulk mode. With recent progress in bottom-up approach, synthesize of long/narrow atomically precise graphene ribbon with well-defined edges has become possible \cite{Ruf,Tej,Bac}. It makes the test of edge-induced magnetism, as well as various properties associated with edge topology, of graphene ribbon possible in the near future. Since all these dynamical effects are related with the bulk chiral states connecting edge states of the system, it is possible for us to generalize our study to other
2D materials with strong edge effect such as transition metal
dichalcogenides \cite{DWu,Zha,Yang} and materials having Kagome \cite%
{Xue} or triangular \cite{Nat,Kur} lattice structure. All of these
are potential directions for further study.

\textit{Acknowledgments.} T. Tan would like to thank Z. Hu for useful discussion. C. Li would like
to thank D. W. Zhai and B. Fu for useful discussions. The work is support by
the University Grants Committee/Research Grant Council of the Hong Kong SAR
(AoE/P-701/20), the HKU Seed Funding for Strategic Interdisciplinary
Research, and the Croucher Senior Research Fellowship.


\begin{thebibliography}{99}







\bibitem{Neto} N. M. R. Peres, F. Guinea, and A. H. Castro Neto, Phys. Rev. B \textbf{73}, 125411 (2006); A. H. Castro Neto, F. Guinea, N. M. R. Peres, K. S.
Novoselov, and A. K. Geim, The electronic properties of graphene, Rev. Mod.
Phys. \textbf{81}, 109, (2009).

\bibitem{Dres} V. Meunier, A. G. Souza Filho, E. B. Barros, and M. S.
Dresselhaus, Physical properties of low-dimensional $sp^{2}$-based carbon
nanostructures, Rev. Mod. Phys. \textbf{88}, 025005 (2016).

\bibitem{Wal} P. R. Wallace, The Band Theory of Graphite, Phys. Rev. \textbf{%
71}, 622 (1947).

\bibitem{Gus} V. P. Gusynin and S. G. Sharapov, Unconventional Integer
Quantum Hall Effect in Graphene,\ Phys. Rev. Lett. \textbf{95}, 146801
(2005).

\bibitem{Novo2} K. S. Novoselov, et. al, Two-dimensional gas of massless
Dirac fermions in graphene,\ Nature \textbf{438}, 197 (2005); M. I.
Katsnelson, K. S. Novoselov, and A. K. Geim, Chiral tunnelling and the Klein
paradox in graphene, Nat. Phys. \textbf{2}, 620 (2006); K. S. Novoselov, et.
al, Room-Temperature Quantum Hall Effect in Graphene, Science \textbf{315},
1379 (2007).

\bibitem{Zhang} Y. B. Zhang, Y.-W. Tan, H. L. Stormer, and P. Kim,
Experimental observation of the quantum Hall effect and Berry's phase in
graphene, Nature \textbf{438}, 201 (2005).


\bibitem{Yao1} Wang Yao, Shengyuan A. Yang, and Qian Niu, Edge States in
Graphene: From Gapped Flat-Band to Gapless Chiral Modes, Phys. Rev. Lett.
\textbf{102}, 096801 (2009).

\bibitem{Del} P. Delplace, D. Ullmo, and G. Montambaux, Zak phase and the
existence of edge states in graphene, Phys. Rev. B \textbf{84}, 195452
(2011).


\bibitem{Nak} K. Nakada and M. Fujita, Edge state in graphene ribbons:
Nanometer size effect and edge shape dependence, Phys. Rev. B \textbf{54},
17954 (1996).

\bibitem{Ryu} Shinsei Ryu and Yasuhiro Hatsugai, Topological Origin of
Zero-Energy Edge States in Particle-Hole Symmetric Systems, Phys. Rev. Lett.
\textbf{89}, 077002 (2002).

\bibitem{Bre} L. Brey and H. A. Fertig, Electronic states of graphene nanoribbons studied with the Dirac equation, Phys. Rev. B \textbf{73}, 235411 (2006).

\bibitem{Mar} I. Martin, Ya. M. Blanter, and A. F. Morpurgo, Topological
Confinement in Bilayer Graphene, Phys. Rev. Lett. \textbf{100}, 036804
(2008).

\bibitem{Yao} D. Xiao, W. Yao, and Q. Niu, Valley-Contrasting Physics in
Graphene: Magnetic Moment and Topological Transport, Phys. Rev. Lett.
\textbf{99}, 236809 (2007); W. Yao, D. Xiao, and Q. Niu, Valley-dependent
optoelectronics from inversion symmetry breaking, Phys. Rev. B \textbf{77},
235406 (2008).

\bibitem{Per} P. M. Perez-Piskunow, G. Usaj, C. A. Balseiro, and L. E. F.
Foa Torres, Floquet chiral edge states in graphene, Phys. Rev. B \textbf{89}%
, 121401(R) (2014).

\bibitem{Zar} M. Zarenia, J. M. Pereira Jr., G. A. Farias, and F. M.
Peeters, Chiral states in bilayer graphene: Magnetic field dependence and
gap opening, Phys. Rev. B \textbf{84}, 125451 (2014).

\bibitem{Ju} L. Ju, et. al, Topological valley transport at bilayer graphene
domain walls, Nature \textbf{520}, 650--655 (2015).

\bibitem{Xi} X. Xi, J. Ma, S. Wan, C.-H. Dong, and X. Sun, Observation of
chiral edge states in gapped nanomechanical graphene. Sci. Adv. \textbf{7},
eabe1398 (2021).


\bibitem{Yaz} O. V. Yazyev, Emergence of magnetism in graphene materials and
nanostructures, Rep. Prog. Phys. \textbf{73}, 056501 (2010).

\bibitem{Son} Y. W. Son, M. L. Cohen, and S. G. Louie, Half-metallic
graphene nanoribbons, Nature \textbf{444}, 347--349 (2006).


\bibitem{Roj} F. Mu\~{n}oz-Rojas, J. Fern\'{a}ndez-Rossier, and J. J.
Palacios, Giant Magnetoresistance in Ultrasmall Graphene Based Devices,
Phys. Rev. Lett. \textbf{102}, 136810 (2009).

\bibitem{Slo} M. Slota, et. al, Magnetic edge states and coherent
manipulation of graphene nanoribbons, Nature \textbf{557}, 691 (2018).

\bibitem{Dut} S. Dutta, A. K. Manna, and S. K. Pati, Intrinsic
Half-Metallicity in Modified Graphene Nanoribbons, Phys. Rev. Lett. \textbf{%
102}, 096601 (2009).

\bibitem{Wim} M. Wimmer, \.{I}. Adagideli, S. Berber, D. Tom\'{a}nek, and K.
Richter, Spin Currents in Rough Graphene Nanoribbons: Universal Fluctuations
and Spin Injection, Phys. Rev. Lett. \textbf{100}, 177207 (2008).




\bibitem{Wan} Z. F. Wang, S. Jin, and F. Liu, Spatially Separated Spin
Carriers in Spin-Semiconducting Graphene Nanoribbons, Phys. Rev. Lett.
\textbf{111}, 096803 (2013).

\bibitem{Li} X. X. Li and J. L. Yang, First-principles design of spintronics
materials, Nat. Sci. Rev. \textbf{3}, 365-361 (2016).

\bibitem{Sah} B. Sahu, H. K. Min, A. H. MacDonald, and S. K. Banerjee,
Energy gaps, magnetism, and electric-field effects in bilayer graphene
nanoribbons, Phys. Rev. B \textbf{78}, 045404 (2008).

\bibitem{Kot} V. N. Kotov, B. Uchoa, V. M. Pereira, F. Guinea, and A. H. C.
Neto, Electron-Electron Interactions in Graphene: Current Status and
Perspectives, Rev. Mod. Phys. \textbf{84}, 1067 (2012).



\bibitem{Lee} G. W. Jeon, K. W. Lee, and C. E. Lee, Layer-selective
half-metallicity in bilayer graphene nanoribbons, Sci. Rep \textbf{5}, 9825
(2015); K. W. Lee and C. E. Lee, Half-metallic quantum valley Hall effect in
biased zigzag-edge bilayer graphene nanoribbons, Phys. Rev. B \textbf{95},
085145 (2017); Topological confinement effects of electron-electron
interactions in biased zigzag-edge bilayer graphene nanoribbons, Phys. Rev.
B \textbf{97}, 115106 (2018).

\bibitem{Sza} K. Sza\l owski, Ferrimagnetic and antiferromagnetic phase in
bilayer graphene nanoflake controlled with external electric fields, Carbon
\textbf{117}, 78-85 (2017).

\bibitem{Yuan} J. Yuan, D.-H. Xu, H. Wang, Y. Zhou, J.-H. Gao, and F.-C.
Zhang, Possible half-metallic phase in bilayer graphene: Calculations based
on mean-field theory applied to a two-layer Hubbard model, Phys. Rev. B
\textbf{88}, 201109(R) (2013).

\bibitem{Ti} Tixuan Tan, Ci Li, and Wang Yao, Edge state in AB-stacked
bilayer graphene and its correspondence with SSH ladder, arXiv: 2109.08462.

\bibitem{Mc} E. McCann and V. I. Fal'ko, Landau-Level Degeneracy and Quantum
Hall Effect in a Graphite Bilayer, Phys. Rev. Lett. \textbf{96}, 086805
(2006);E. McCann and M. Koshino, The electronic properties of bilayer
graphene, Rep. Prog. Phys. \textbf{76}, 056503 (2013).

\bibitem{Kuz} A. B. Kuzmenko, I. Crassee, D. van der Marel, P. Blake, and K.
S. Novoselov, Determination of the gate-tunable band gap and tight-binding
parameters in bilayer grapheneusing infrared spectroscopy, Phys. Rev. B
\textbf{80}, 165406 (2009).

\bibitem{Hartmann} T. Hartmann, F. Keck, H. J. Korsch and S. Mossmann,
Dynamics of Bloch oscillations, New. J. Phys \textbf{6}, 2 (2004).

\bibitem{Viktor} V. Krueckl, and K. Richter, Bloch-Zener oscillations in
graphene and topological insulators, Phys. Rev. B \textbf{85}, 115433 (2012).

\bibitem{Liu} Xiong-Jun Liu, K. T. Law, T. K. Ng, and Patrick A. Lee, Detecting Topological Phases in Cold Atoms, Phys. Rev. Lett. \textbf{111}, 120402 (2013).

\bibitem{Bloch} M. Atala, M. Aidelsburger, J. T. Barreiro, Dm. Abanin, T. Kitagawa, E. Demler, and I. Bloch, Direct measurement of the Zak phase in topological Bloch bands, Nat. Phys \textbf{9}, 795-800 (2013).

\bibitem{Chunyan} C. Li, W. Zhang, Y. V. Kartashov, D. V. Skryabin, and F.
Ye, Bloch oscillations of topological edge modes, Phys. Rev. A \textbf{99},
053814 (2019).

\bibitem{Lib} M. D. Liberto, N. Goldman, and G. Palumbo, Non-Abelian Bloch oscillations in higher-order topological insulators. Nat. Commun \textbf{11}, 5942 (2020).

\bibitem{Note4} One can also understand this fact from the well-known
formula describing Landau-Zener tunneling \cite{Amit}: $P=e^{-2\pi \Delta
^{2}\tau }.$ Where $P$ is the transition probability between two bands, $%
\Delta $ is approximately the gap between two bands, $\frac{1}{\tau }$ is
approximately the asymptotic slope of the band. Since $\tau $ is infinitely
large for our topological flat edge bands, while $\Delta $ is finitely
small, then the transition is infinitely suppressed. When near the Dirac
point, two edge bands become dispersive and their wavefunctions have
appreciable overlap. However, a gap around $30$ $\mathrm{meV}$ is opened
even for zero bias situation, as shown in Fig. \ref{fig1}, which makes the
interband transition near the Dirac point still weak. This leads the Bloch
oscillation in our system to distinct from those discussed in the
literature, such as Ref. \cite{Viktor,Chunyan}. There the dispersive
topological edge bands inevitably need to take Bloch-Zener transition into
consideration. It is also distinct from the Bloch oscillation in bulk
bilayer graphene. In that condition the bulk bands which are responsible for
the Bloch oscillation are highly dispersive, facilitating the transition
between nearby bands via Landau-Zener tunneling.

\bibitem{Amit} A. Dutta, G. Aeppli, B. K. Chakarabarti, U. Divakaran, T. F.
Rosenbaum, and D. Sen, \textit{Quantum Phase Transitions in Transverse Field
Spin Models: From Statistical Physics to Quantum Information} (Cambridge
University Press, Delhi, 2015).

\bibitem{Supp} See Supplementary for Bloch oscillation behavior with smaller
bias, details on self-consistent mean-field calculation.

\bibitem{Note3} Difference in the small/large deviation is the magnitude of
the electric field. Here, the important time scale are the momentum
relaxation time $\tau_p$ and the spin relaxation time $\tau_s$. In graphene
nanoribbons with magnetism on the edge, the typical $\tau_p$ is of order of $%
1\ \mathrm{ps}$ \cite{Ertler}, and $\tau_s$ can be the order of $10\sim 100\
\mathrm{ps}$ \cite{Kat1,Kat2}. If the field is too small and relaxation sets
in before the deviation is large enough, it corresponds to condition $(1)$
of Fig. \ref{fig3}(b). On the contrary, if the field is large enough to take
electrons far away from the initial configuration in the momentum space
within $\tau_p$, then it corresponds to condition $(3)$ of Fig. \ref{fig3}%
(b), which makes the ground-state exchange, i.e., spin filp possible under
the spin relaxation. Therefore, it is easy to give an estimated thershold
electric field for the switching described in the text to appear: $%
E_{threshold}\sim\frac{\pi\hbar}{ae\tau_p}\sim 10^{-2}\ \mathrm{V/nm}$,
where $a=0.246\ \mathrm{nm}$ is the lattice constant of MLG.

\bibitem{Ertler} C. Ertler, S. Konschuh, M. Gmitra, J. Fabian, Electron spin
relaxation in graphene: The role of the substrate, Phys. Rev. B \textbf{80},
041405(R) (2009).

\bibitem{Kat1} V. K. Dugaev and M. I. Katsnelson, Spin relaxation related to
edge scattering in graphene, Phys. Rev. B \textbf{90}, 035408 (2014).

\bibitem{Kat2} M. I. Katsnelson, \textit{The Physics of Graphene, 2nd ed}
(Cambridge University Press, Cambridge, 2020).

\bibitem{Note2} The SP is expressed by the magnetic moment $M_{i}\equiv
\left\langle n_{i,\uparrow }\right\rangle -\left\langle n_{i,\downarrow
}\right\rangle $ at site $i$, the average spin-up/down electron population $%
\left\langle n_{i,\uparrow /\downarrow }\right\rangle $ is determined by the
self-consistent mean-field calculation (see \cite{Supp} for details). In
this letter, we take $t=3.16$\textrm{eV} \cite{Kuz}.

\bibitem{Tej} L. Brey, P. Seneor, and A. Tejeda, Chapter \textbf{2}, \textit{Graphene Nanoribbons} (IOP Publishing Ltd 2020).

\bibitem{Bac} C. Backes, et. al, Production and processing of graphene and related materials, 2D Mater. \textbf{7}, 022001 (2020).

\bibitem{Ruf} P. Ruffieux, et. al, On-surface synthesis of graphene
nanoribbons with zigzag edge topology, Nature \textbf{531}, 489-492 (2016).

\bibitem{DWu} D. Wu, et. al, Uncovering edge states and electrical
inhomogeneity in MoS2 field-effect transistors, Proc. Natl. Acad. Sci.
U.S.A. \textbf{113}, 8583-8588 (2016).

\bibitem{Zha} C. Zhang, et. al, Visualizing band offsets and edge states in
bilayer-monolayer transition metal dichalcogenides lateral heterojunction.
Nat. Commun. \textbf{7}, 10349 (2016).

\bibitem{Yang} G. Yang, Y. Shao, J. Niu, et. al, Possible Luttinger liquid
behavior of edge transport in monolayer transition metal dichalcogenide
crystals. Nat. Commun. \textbf{11}, 659 (2020).



\bibitem{Xue} H. R. Xue, Y. H. Yang, F. Gao, Y. D. Chong, and B. L. Zhang,
Acoustic higher-order topological insulator on a kagome lattice. Nat. Mat.
\textbf{18}, 108-112 (2019).

\bibitem{Nat} P. Nataf, M. Lajk{\'o}, A. Wietek, K. Penc, F. Mila, and A. M.
L{\"a}uchli, Chiral Spin Liquids in Triangular-Lattice $SU(N)$ Fermionic
Mott Insulators with Artificial Gauge Fields, Phys. Rev. Lett. \textbf{117},
167202 (2016).

\bibitem{Kur} T. Kurumaji, et. al., Skyrmion lattice with a giant
topological Hall effect in a frustrated triangular-lattice magnet, Science
\textbf{30}, 914-918, (2019).
\end{thebibliography}
\end{document}